%% file: ms.tex
\documentclass[preprint2]{aastex}
\begin{document}
\title{Microlensing of Relativistic Knots in the Quasar HE1104$-$1805}

\shorttitle{Microlensing of HE1104$-$1805} 
\shortauthors{OGLE Collaboration} 
\author{Paul L. Schechter\altaffilmark{1,2},
A.~Udalski\altaffilmark{3},
M.~Szyma{\'n}ski\altaffilmark{3},
M.~Kubiak\altaffilmark{3},
G.~Pietrzy{\'n}ski\altaffilmark{3,4},
I.~Soszy{\'n}ski\altaffilmark{3},
P.~Wo{\'z}niak\altaffilmark{5},
K.~{\.Z}ebru{\'n}\altaffilmark{3},
O.~Szewczyk\altaffilmark{3},
and {\L}. Wyrzykowski\altaffilmark{3}}
\altaffiltext{1}{Massachusetts Institute of Technology, Cambridge, MA~02139-4307, USA
{\tt schech@achernar.mit.edu}
}
\altaffiltext{2}{Institute for Advanced Study, Princeton, NJ~08540-0631, USA
}
\altaffiltext{3}{Warsaw University Observatory, Al.~Ujazdowskie 4, 00-478 Warszawa, Poland
{\tt (udalski,msz,mk,pietrzyn,
soszynsk,zebrun,szewczyk,wyrzykow)@astrouw.edu.pl}
}
\altaffiltext{4}{Universidad de Concepci\'on, Departamento de Fisica
Casilla, 160-C, Concepci\'on, Chile
{\tt pietrzyn@hubble.cfm.udec.cl}
}
\altaffiltext{5}{Los Alamos National Laboratory, MS-D436, Los Alamos,
NM~87545, USA
{\tt wozniak@lanl.gov}
}

\begin{abstract} 
We present 3 years of photometry of the ``Double Hamburger'' lensed
quasar, HE1104$-$1805, obtained on 102 separate nights using the OGLE
1.3-m telescope.  Both the A and B images show variations, but with
substantial differences in the lightcurves at all time delays.  At the
$310^{\rm d}$ delay reported by Wisotzki and collaborators the difference
lightcurve has an rms amplitude of 0.060 mag.  The structure functions
for the A and B images are quite different, with image A more than
twice as variable as image B (a factor of 4 in structure function) on
timescales of less than a month.  Adopting microlensing as a working
hypothesis for the uncorrelated variability, the short timescale
argues for the relativistic motion of one or more components of the
source.  We argue that the small amplitude of the fluctuations is due
to the finite size of the source with respect to the microlenses.
\end{abstract} 

\section{INTRODUCTION}

Two very different physical processes contribute to the observed
photometric variability of gravitationally lensed quasars: the
intrinsic variabilty of the quasar itself and propagation effects
along the line of sight.  Chief among the latter is microlensing by
the stellar mass objects in the intervening lens (Chang and Refsdal
1979; Paczy{\'n}ski 1986).  The combination of intrinsic and microlensing
variations represents an embarassment of riches.  For the purpose of
measuring lens time delays (using the correlated intrinsic variabilty
of the quasar images) uncorrelated microlensing variations are an
additional source of noise.  Conversely, the intrinsic variation of
the quasar produces correlated noise in the uncorrelated microlensing
signals.

Time delays have been measured for nearly a dozen systems, and in most
cases microlensing appears not to have presented a serious problem
(e.g. Kundi{\'c} et al. 1997; Schechter et al. 1997).  Dramatic
microlensing variations have been observed in the system 2237+0305
(Corrigan et al. 1991; Wo{\'z}niak et al. 2000) but on a timescale
(months) which is very much longer than the predicted delays (hours).

There have, however, been instances in which microlensing and time
delay measurements have interfered with each other.  In the case of
0957+561, the two images have long timescale (1000$^{\rm d}$) variations
(Refsdal et al. 2000), which bias the inferred time delay.  Burud et
al. (2000) report uncorrelated variations over timescales of several
months in their study of B1600+434.  Examination of their Figure 3
shows apparent uncorrelated variations on timescales of days.
Uncorrelated variations are also reported by Hjorth et al.\ (2002) in
their study of RXJ0911+0551.

We report here the results of an unsuccessful program to measure the
time delay of the doubly imaged quasar HE1104$-$1805 (Wisotzki et
al. 1993).  In three years' monitoring with the Optical Gravitational
Lensing Experiment (OGLE) 1.3m telescope at Las Campanas we see
uncorrelated variations in the A and B images, which we interpret as
the result of microlensing.

In \S 2 we describe the observations and initial reductions.  In
\S 3 we compare light curves for the two quasar images, A and B,
using the time tested chi-by-eye technique and a less subjective
method.  Our data fail to produce a satisfactory time delay.  In
\S 4 we derive structure functions separately for the A and B
images.  Adopting the time delay measured by Gil-Merino et al.\ (2002),
we determine a structure function for the microlensing from the
difference between the A and B images.  In \S 5 we explore several
alternative interpretations of the structure functions for the two
quasar images.
\section{OBSERVATIONS}

Observations of HE1104$-$1805 were carried over a three year period as a
sub-project of the second phase of the OGLE microlensing search
(Udalski, Kubiak and Szyma{\'n}ski 1997). The 1.3-m Warsaw telescope
at the Las Campanas Observatory, Chile (operated by the Carnegie
Institution of Washington) was equipped with the ``first generation"
camera, incorporating a SITe ${2048\times2048}$ CCD detector working
in still-frame mode.  The pixel size was 24~$\mu$m, giving a scale of
0\farcs417/pixel. Observations were performed in the ``medium''
readout mode of the CCD detector, with a gain 7.1~e$^-$/ADU and
read noise of about 6.3~e$^-$. Details of the instrumentation setup
can be found in Udalski, Kubiak and Szyma{\'n}ski (1997).

Our plan was to obtain data three times per month, at moon phase
$-11$, $-1$ and $+8$ days, weather permitting.  If not, data was
obtained on the first good night following the planned observation.
Typically two 10 minute exposures were taken with a standard $V$-band
filter. The data were reduced automatically at the telescope using the
OGLE software pipeline.  Photometry of the two quasar components, A
and B, and for three reference stars, CA, CB and CC was derived with
the {\sc DoPhot} photometry program (Schechter, Mateo and Saha 1993).
The third of these reference stars proved to be variable.  Stars CA
and CB are --3\farcs2 and 31\farcs9 to the East and 15\farcs1 and
5\farcs3 to the South, respectively from the brighter quasar
component, A.  All magnitudes are referred to the average of these two
stars.  No attempt was made to correct either for color terms or for
an airmass/color crossterm.  Figure~1 presents a $120''\times 120''$
subraster of a $V$-band frame of HE1104$-$1805 showing both quasar
images, A and B, and both reference stars, CA and CB.

The complete record of observations is given in Table 1.  Three
exposures which gave very different magnitudes from the exposure
immediately preceeding or following and from magnitudes obtained
earlier or later in the month are flagged and are not used in the
discussion that follows.  Data taken on the same night was averaged
and was assigned the average time of observation.  Nightly averaged
photometry for quasar images A and B and for comparison star CA is
shown in Figure~2.  Note that there is no additional information in
star CB, as its variations mirror those of CA, but with the opposite
sign. We do not plot formal error bars, as night-to-night variations
are always larger than these.
\section{TIME DELAY}

Wisotzki and collaborators (Wisotzki et al. 1998; Gil-Merino et
al. 2002) have measured a time delay of $310 \pm 19^{\rm d}$ for
HE1104$-$1805, with the B image leading the A image.  In Figure~3 we
plot our photometry (averged over each night, typically two exposures)
for image B at the corresponding time for image A, $310^{\rm d}$ later than
actually observed.  The photometry for image A is shown as actually
observed.

\subsection{Difference Lightcurve}

As we are looking at a single quasar along lines of sight that differ
by only a few arcseconds, the instrinsic component of the A and B
variability ought to be the same after compensation for the time
delay.  A lagged difference lightcurve should show only the effects
of extrinsic processes, e.g. microlensing.  The price we pay in so
doing is that the difference lightcurve includes extrinsinc variations
(and noise) from both images.

In computing a lagged difference, we chose to interpolate on the B
lightcurve, which is considerably smoother.  Our approach was to use
all data taken within 20 days of the desired time.  We fit a straight
line (quadratic) if the desired time was straddled by two (three or
more) observations; if not we took a straight average.

Figure 3 shows a difference lightcurve, obtained by interpolating on
image B as described above.  While there are modest similarities in
the two lightcurves, the differences are substantial.  

It should be noted that our single inital observation of image B in
August 1997 (HJD 2450666) yields 5 points at the extreme left of the
difference curve.  We choose to keep these points because of the extra
baseline they gain us.  The advisability of this rests heavily on the
assumption that B varies slowly.

\subsection{Time Delay}

Over the course of the three years spent observing HE1104$-$1805, with
the acquisition of every new data point, we attempted to determine a
time delay using the time honored ``chi-by-eye'' approach.  At no
point were we able to persuade ourselves that we had a robust
determination of the time delay.  In particular there are many
significant features in the A lightcurve for which the B lightcurve
has no counterpart.  A variety of sophisticated techniques have been
developed to extract time delays (see Gil-Merino et al. 2002 and
references cited therein), but should one believe a time delay that
fails the chi-by-eye technique?

We subscribe to Lord Rutherford's dictum (Bailey 1967) -- ``If your
experiment needs statistics, you ought to have done a better
experiment.''  But since the community standard would seem to be that
one must carry out an ``objective'' test, we have done so.  As the B
image seems to vary less than the A image, we interpolate over the B
values as described above to obtain a prediction for A at a trial
value of the time delay.  We then compute the rms deviation between
the observed magnitude for a wide range of time delays, plotted in
Figure 4.  The rms scatter at the published time delay is 0.060 mag,
with 51 overlapping points.  There is a minimum of 0.054 at $145^{\rm d}$
with 56 overlapping points.  Without knowing the statistics of the
uncorrelated features we cannot argue whether either of these is to be
preferred over yet some other value.

We admit defeat.  We cannot measure the time delay using our data and,
for the purpose of the ensuing discussion, we use the value of
Gil-Merino et al.  If the variations intrinsic to HE1104$-$1805 were
large, the error in our adopted delay would would introduce a spurious
microlensing signal.  But had the intrinsic variations been large, we
would have been able to measure the delay.
\section{VARIABILITY IN THE A AND B IMAGES}

\subsection{Structure Functions for A and B}

It is our subjective impression that the A lightcurve in Figure 2 is
much more variable than the B lightcurve.  This can be made
quantitative by computing structure functions for images A and B, 
\begin{equation}
s(\tau) = \Big<[(m(t+\tau) - m(t)]^2\Big>
\quad ,
\end{equation}
where angle brackets denote an average over time.  Figure 5 shows
these functions computed for images A and B, in 9 day bins.  As in the
previous section we use the nightly averages to compute individual
values of $m(t)$.  The first bin includes lags $0 < \tau < 4\fd5$, the
second lags $4.5 < \tau < 13\fd5$, and so forth.  The structure
function for image A shows considerably more variability than that for
image B.  For lags shorter than 40 days it is a factor of four
larger, corresponding to a factor of two more variability.  We compute
structure functions for the difference between comparison stars A and
B, and find a roughly constant amplitude of 0.0002, corresponding to
photometric errors of 0.014 mag.  If measurement errors make a similar
contribution to image B, the structure function for image A at 9 days
is a factor of eight larger than for B.

\subsection{Working Hypotheses}

The A image is brighter than the B image, but not so bright that the
images are anywhere near saturation.  We can think of no reason why
the measurement errors for A should be larger than for B.  As we
have made no attempt to account for the difference in color between
the comparison stars and the quasar images, one might expect
photometric errors to show a seasonal correlation (and hence an
increasing structure function) depending upon whether the object was
observed at low or high airmass.  But the variations seen in both the A
and B images are so large that we take them to be the result of
physical processes beyond the Earth's atmosphere.

We take the difference in the amplitudes of the two structure
functions to be significant, indicating that whatever physical
processes are at work, they affect the images differently.  Were we
working at radio wavelengths, scintillation arising in the intervening
galaxy would  need to be considered, as it was by Koopmans and de
Bruyn (2000) in the case of B1600+434.  For lack of a better
explanation (and perhaps a lack of imagination), we adopt, as our
working hypothesis, that the fluctuations in image A are due largely
to microlensing by stars in the lensing galaxy.

Some variation is expected due to the intrinsic variability of the
quasar.  The structure functions for quasars typically have amplitudes
of (0.1 mag)$^2$, with time constants of order one year (Cristiani et
al. 1996).  There are, of course, very substantial variations among
quasars, with some showing almost no variability and others showing
intraday variability.  The lightcurves and structure functions for A
and B look so different that we take the variations in image B to be
largely instrinsic, consistent with our expectations for typical
quasars.  A smaller microlensing amplitude might naively be expected
because the surface brightness of the lensing galaxy is much lower at
the position of image B and because we expect a continuous dark matter
component to be a larger fraction of the mass.  While the effects of
dark matter are not quite so simple (Schechter and Wambsganss 2002;
see below), it is nonetheless the case that microlensing should indeed
affect A more than B.

\subsection{Difference Structure Function}

Figure 6 shows the structure function obtained for the difference
lightcurve presented in Figure~3.  We take its large fluctuations to be
the accidental beating of features in lightcurve A against features in
lightcurve B.  For the purpose of comparison with microlensing
simultations, we will want a charactistic timescale for the difference
lightcurve.  While the curve is exceedingly noisy, we estimate the
time for our structure function to rise to one quarter of its
asymptotic value, $T_{1/4}$, to be $\sim 20^{\rm d}$.
\section{INTERPRETATION}

\subsection{Microlensing at High Optical Depth}
High optical depth microlensing differs qualitatively from its low
optical depth counterpart.  A great many microimages contribute to the
observed flux of macroimages like A and B in HE1104-1805
(Pacy\'nski 1986).  Their combined flux may be efficiently simulated
by tracing rays back from the observer to the source plane (Kayser et
al. 1986; Wambsganss 1992; Lewis and Irwin 1995), producing a
magnification map.  As a source moves with respect to the lensing
galaxy, it cuts across this magnification map, producing a simulated
lightcurve.

At low optical depth such lightcurves have long, relatively constant
intervals, punctuated by occasional large fluctuations.  At high
optical depth the lightcurves are more uniform in time, but there are
still large amplitude ``events'' caused by the creation and
annihilation of pairs of microimages as the source crosses
``caustics,'' and by close encounters with
``cusps'' in the magnification map.  To first order the structure
functions for such simulated lightcurves have a rise time whose scale
is governed by the size the Einstein rings of the microlenses and
whose amplitude depends upon the surface density of microlenses and
the local shear (Lewis and Irwin 1995; 1996).

A minimum of three parameters is needed to characterize the
microlensing of a macroimage: a dimensionless surface density of
microlenses, $\kappa_*$, a dimensionless surface density of a smooth
``dark'' component, $\kappa_c$, and a dimensionless tidal stretching or
shear, $\gamma$.  For a mulitply imaged quasar, the image positions
(and sometimes magnifications) give a model that constrains $\gamma$
and $\kappa_{tot} \equiv \kappa_c + \kappa_*$ at the position of each
image.  Paczy\'nski (1986) shows that the relevant model space
collapses to an equivalent two dimensional space spanned by an
effective dimensionless surface density, $\kappa^{eff}_* =
\kappa_*/(1-\kappa_c)$ and an effective shear, $\gamma^{eff} =
\gamma/(1-\kappa_c)$, but with the magnifications larger by
$1/(1-\kappa_c)^2$ than those computed from the effective values.

The amplitudes of the microlensing fluctuations go to zero in the
limits both of unit magnification (no microlenses) and in the case
very high magnification (when fluctuations in the number
of microimages become negligible).  Taking the effective
magnification to be given by $\mu^{eff} = 1/[(1-\kappa_*^{eff})^2 -
(\gamma^{eff})^2]$, the largest microlensing fluctuations occur when
$\mu^{eff} \sim 3$ (Granot et al. 2002).  Macroimages that are
saddlepoints of the arrival time surface (those with $1 - \kappa -
\gamma < 0$) show larger fluctuations than those that are minima.
The fluctuation amplitudes for minima appear to peak at roughly one
magnitude, but those for saddlepoints can be somewhat larger
(Schechter and Wambsganss 2002).  The magnification histograms for
both minima and maxima are bimodal in the vicinity of $\mu^{eff} \sim
3$.  A histogram computed from the observed difference lightcurve for
HE1104-1805 is shown in Figure 7.

\subsection{Macromodel for HE1104$-$1805}

We have used the Leh\'ar et al. (2000) HST positions for the components
of HE1104$-$1805 and the lensing galaxy, and an emission line flux ratio $A/B
= 2.8$ (Wisotzki 1993) to model the lensing potential as a singular
isothermal sphere in the presence of external shear.  The
dimensionless surface density, shear and magnification for the A and B
images are given in Table 2, with negative magnification indicating a
saddlepoint.  The shear is sufficiently large in this system that it
seems unlikely to be the result of flattening of the lensing
potential.  Fitting models which had steeper (shallower) potentials
would give lower (higher) surface densities and lower (higher)
magnifications, but for reasonable ranges the conclusions of the
subsequent sections are essentially unchanged.

\subsection{Fluctuation Amplitude: Image A}

The structure function computed from a lightcurve (or its two
dimensional counterpart, computed from a magnification map) will be
characterized by a dimensionless amplitude and a time scale (or length
scale).  At large temporal (or spatial) separations, we expect
the amplitude to saturate at some asymptotic value.  The amplitude is
governed by two dimensionless parameters: the ratio of the size of the
source to the size of a microlens (its Einstein ring radius), and the
fraction of the total surface density in microlenses, as opposed to a
smoothly distributed component.  The time scale is set by dividing
the size of the Einstein ring by the velocity of the source relative
to the microlenses.

The size of the Einstein ring clearly affects both the fluctuation
amplitude and the timescale.  Here we adopt an idealized model which
makes it particularly easy to separate the effects of source size from
those of dilution by smoothly distributed matter.  We take the source
to have two components, one very much smaller than the microlenses and
one very much bigger.  In such a scenario, the amplitude of the
fluctuations is smaller by the ratio of the flux in the smaller
component to the total flux.  We draw tentative conclusions using
this model and then consider how relaxing our assumptions might
affect them.

\subsubsection{Dark Matter Dilution}

The observed distribution for the shifted magnitude differences
between the A and B images of HE1104$-$1805 has an rms scatter of 0.060
magnitudes.  Our working hypothesis is that these result from
microlensing fluctuations in image A.  Lewis and Irwin (1995, 1996) present
microlensing results for a point source with $(\kappa_*,\gamma) =
(0.65,0.50)$ -- approximately correct for image A if
$\kappa_*/\kappa_{tot} = 1$.  They predict fluctutations of order 1
magnitude.  An extended source would give smaller fluctuations.
Alternatively, one might suppose that increasing the smooth component,
$\kappa_c$, at the expense of the microlensing component, $\kappa_*$,
would likewise produce smaller fluctuations.

We therefore prevailed upon J. Wambsganss to carry out a series of
microlensing simulations for $(\kappa_{tot},\gamma) = (0.60,0.50)$,
decreasing $\kappa_*/\kappa_{tot}$ from 1 to 0.  We were flabbergasted
to discover that the fluctuations increased rather than decreased, at
least at first (Schechter and Wambsganss 2002).  Eventually the
fluctuations get small, but only when $\kappa_*/\kappa_{tot} << 1$.
At such levels of dilution, the magnification histogram is exceedingly
asymmetric, with a relatively narrow spike at the expected
magnfication and a tail roughly 2 magnitudes long toward large
demagnifications.  This is not the character of the fluctuations
observed in Figure 7.  Moreover this would require an extraordinarily
low mass-to-light ratio for the stellar component of the lensing
galaxy.  We conclude that a model in which the small observed
fluctuations are due to dilution by dark matter is not viable.

But we {\it do} expect some dark matter dilution.  
In their study of the lensing galaxy in the system 0047$-$281, Koopmans and 
Treu (2002) find the stellar mass fraction interior to the Einstein
ring to be $0.58 \pm 0.04$.  It seems reasonable to take it to be a 
factor of $\sim 2$ smaller {\it at} the Einstein ring.  If HE1104$-$1805 were
similar, it would imply $(\kappa_*/\kappa_{tot})_A = 0.35$.

\subsubsection{A Source with a Big and a Small Component}

For our idealized model, the fluctuations are diluted by the ratio of
the flux in the compact component to the total flux.  In Wambsganss'
simulations the rms fluctuations vary from 0.69 mag for
$\kappa_*/\kappa_{tot} = 1$ through 1.21 mag for
$\kappa_*/\kappa_{tot} = 0.16$, beyond which the fluctuations begin to
decrease.  At $\kappa_*/\kappa_{tot} = 0.35$ the rms fluctuations are
about 1 magnitude.  A microlensed hot spot would therefore need $\sim$
7\% of the total flux to produce the observed lensing histogram.

Interestingly, the magnification histograms for $0.15
<\kappa_*/\kappa_{tot} < 0.50$ bifurcate into two peaks, separated by
roughly 1.6 magnitudes (cf. Figure 3 of Schechter and Wambsganss 2002).
A hotspot with 7\% of the total flux would produce two fluctuation
peaks separated by 0.10 mag.  Aided by theory, one can imagine such a
bifurcation in the observed histogram for HE1104$-$1805, Figure 7.

\subsection{Fluctuation Amplitude: Image B}

Figure 5 shows considerably less fluctuation in image B than in image
A.  This might be expected both because the stellar surface density is
lower at B than at A, and because the dark matter fraction is higher
at B.  Substituting dark matter for microlenses might increase the
microlensing fluctuation for high magnification images, especially
saddlepoints, but it does not for lower magnification minima.

Our isothermal model gives us the total surface density at A and B,
with $\kappa_{tot,B}/\kappa_{tot,A} = 0.52$.  Leh\'ar et al. (2000)
measure the effective radius for the lensing galaxy, $r_e =
0\farcs73$.  Images A and B are 1\farcs10 and 2\farcs12, respectively,
from the lensing galaxy.  For constant M/L we then have
$\kappa_{*,B}/\kappa_{*,A} = 0.22$.  We therefore expect
$(\kappa_*/\kappa_{tot})_B = 0.42 (\kappa_*/\kappa_{tot})_A $.

If we take $(\kappa_*/\kappa_{tot})_A = 1$, we get $\kappa_{*,B} =
0.140$ and $\kappa_{c,B} = 0.204$. These give $\kappa^{eff}_* = 0.176
$ and $\gamma^{eff} = 0.263 $.  Under these circumstances, we would
expect, according to Lewis and Irwin (1996), microlensing fluctuations
in image B to be roughly half as large as in image A, which is
marginally consistent with the structure functions of Figure 5, though
perhaps not at the shortest lags.

If we take $(\kappa_*/\kappa_{tot})_A = 0.35$, the fluctuations for
image A increase and the fluctuations for image B decrease.  In this
case most of the fluctuation in image B would be due to intrinsic
variations in the quasar.

In any event, the character of the microlensing light curve predicted
for B is quite different than that predicted for A.  At the lower
surface density of image B we expect infrequent but high magnification
events (and pairs of events), with the spacing between events large
compared to their rise times and durations.  Substituting dark matter
for microlenses makes such events yet more infrequent.  We may not have
sampled image B long enough to see any of its occasional fluctuations.

\subsection{Tentative Conclusions}

While the small amplitude of the fluctuations is primarily due to
finite source size effects, the above arguments would seem to favor
some dilution of the stellar component with dark matter at both A and
B for the following reasons: a) it minimizes the flux of the hotspot
b) it reduces the predicted amplitude of microlensing fluctuations in
image B and c) it produces a bimodal magnification histogram for image
A. 

\subsection{Timescale}

To first approximation, simulated microlensing structure functions
look like the voltage across the capacitor in an RC circuit, rising
and then asymptotically approaching a constant value.  Lewis and Irwin
(1996) add an additional parameter that allows for more protracted
rises, but for the present discussion it suffices to think in terms of
their $T_{1/2}$, the time it takes for their structure function
to reach half its asymptotic value.  Lewis and Irwin use a variant of
the structure function which computes the sum of absolute values of
differences rather than the sum of squares in equation (1), so their
$T_{1/2}$ corresponds roughly to our $T_{1/4}$.  For a
range of cases with $\kappa_* = \gamma$ and no smooth component (as
would be appropriate for a singular isothermal sphere), they find
$T_{1/2} \sim 0.33$, (measured in Einstein ring radii per unit
time) for source trajectories that cross the shear direction at a
$45\degr$ angle to the shear.

The radius of the Einstein ring, projected back onto the source plane,
is given by 
\begin{equation}
R_E = {\left[{4 G M \over c^2} {D_{LS} D_S \over D_L} 
\right]}^{\onehalf}
\end{equation}
where $D_L$, $D_S$, $D_{LS}$ and are angular diameter distances to the
lens, the source and from the lens to the source, respectively.  The
redshifts for the lens and quasar are 0.729 and 2.319, respectively
(Lidman et al. 2000),
giving $D_L H_0/c = 0.350$, $D_S H_0 / c = 0.395 $ and $D_{LS} H_0/c =
0.212$ for $(\Omega_m, \Omega_\Lambda) = (0.3,0.7)$.
We therefore have 
\begin{equation}
R_E = 4.31 \times 10^{16} {\rm cm}
{\left({M \over M_\sun}\right)}^{\onehalf}
{\left({70 {\rm km/s/Mpc} \over H_0}\right)}^{\onehalf} \quad {\rm and}
\end{equation}
\begin{equation}
T_{1/2} \sim 5\fd55
{\left({c \over v}\right)}
{\left({M \over M_\sun}\right)}^{\onehalf}
{\left({70 {\rm km/s/Mpc} \over H_0}\right)}^{\onehalf} \quad,
\end{equation}
where our decision to normalize the source velocity by the speed of
light reflects the remarkably short timescale for the observed
fluctuations.  Our observed $T_{1/4} \sim 20^{\rm d}$ timescale
implies a source velocity of $\sim 0.25 c$ for solar mass microlenses.

\subsection{Alternative Models and Interpretations}

\subsubsection{Intermediate Sized Optical Continuum}

Our working model takes the source to have two components, one very
much smaller than the microlenses and one very much bigger.  The range
of possible alternative models for the surface brightness distribution
of the source is limited only by imagination.  Source models favored
in the microlensing literature include Gaussians and uniform disks as
well as more physically motivated models (Agol and Krolik 1999).  In
the case of HE1104$-$1805, source velocities approaching the speed of
light are indicated.  It seems unlikely that the entire optical
continuum region would be move, as a unit, with such a velocity.

There is evidence, however, that the continuum source cannot be too
much larger than the microlenses.  In their discovery paper, Wisotzki
et al.\ (1993) note that while the emission line flux ratio is
constant at a factor of 2.8, the continuum flux ratio increases to the
blue, rising to a factor $\sim 6$ in the B filter.  They argue that
the continuum source is likely to be more compact than the broad line
region, and that the continuum is microlensed.  In this case we would
be underestimating, by roughly a factor of 2, the fractional
contribution of the fast moving region to the unlensed flux.

\subsubsection{Nanolensing by $10^{-5} M_\sun$ Planets}

An alternative explanation that would obviate the need for
relativistic velocities invokes microlenses with planetary masses, of
order $10^{-5} M_\sun$ (Schild 1996 and references cited therein).
The velocity of the source relative to the lensing system would then
be of order $10^{-3}c$.  Such ``nanolenses'' impose correspondingly
tighter constraints on the size of the compact part of the source.
Wyithe and Loeb (2002) therefore argue that hotspots are still
needed in such a model: the accretion disk would be larger than the
nanolenses but the hotspots might be smaller.

An additional difficulty associated with this hypothesis is that
microlensing by stars might interfere with nanolensing.  The
anomalously large fluctuations found by Schechter and Wambsganss
(2002) depend critically on the lensing potential being 
smooth on larger scales.  If the macro-image is broken up into 
micro-images, nanolenses will operate separately on each of these 
to smaller effect.

\subsubsection{Coldspots}

Wyithe and Loeb (2002) argue that short timescale fluctuations
observed by Hjorth et al. (2002) and Burud et al. (2002) might be due
the the microlensing of broad absorption line clouds shadowing a
quasar accretion disk.  None of the figures they present for these
models look much like our data: either the amplitudes are too small or
the timescales are too long.

\subsubsection{Multiple Hotspots}

Interestingly, Wyithe and Loeb also simulate hotspots that arise
within the quasar's accretion disk, as proposed by Gould and
Miralda-Escud{\'e} (1997).  These have orbital velocities $\sim 0.2 c$.
Their model with only a single hotspot shows a strong periodicity and
the classic M-shaped profile associated with the creation and
annihilation of a single pair of micro-images.  While we see no such a
periodic signal in our data, it would be premature to rule out a
single, accretion disk hotspot on the basis of a single set of model
parameters.  A relativistic hotspot in a jet would likewise not be
excluded by their model.

Interestingly, the lightcurve for their model with 100 hotspots shows
considerably more resemblance to the difference lightcurve for
HE1104$-$1805, both in amplitude and scale.  Moreover their parameters,
$\kappa_{tot} = \gamma = 0.54$ and $\kappa_* = 0.08$, are quite
similar to those we adopt for the B image of HE1104$-$1805.  They
produce a saddlepoint with a magnification $\mu = -12.5$ and microlens
fraction of 15\%, almost ideal for the purpose of maximizing
microlensing fluctuations.
\section{SUMMARY}
Three years of observations of the two lensed quasar images of
HE1104$-$1805 show variations in the A and B images that are
uncorrelated, with V amplitudes of $\sim 0.060$ mag. The A image
exhibits considerably more variability, on a timescale $\lesssim$ 1
month, while the fluctuations in the B image are consistent with our
expectations for variations intrinsic to the quasar.

On the hypothesis that the fluctuations are due to microlensing by
solar mass stars, the implied source velocity is $\sim 0.25 c$.  For
reasonable assumptions regarding the ratio of dark to microlensing
matter and on the assumption that only a single hotspot is
contributing to the fluctuations, the hotspot contributes $\sim 7$\%
of the continuum flux.  A multiple hotspot model presented (and
rejected) by Wyithe and Loeb (2002) also seems viable, perhaps even
preferable, in the present case.
\acknowledgements We thank Professor B.\ Paczy{\'n}ski for his
support and Professor J.\ Wambsganss for his counsel.  We gratefully
acknowledge the support of the US NSF through grants AST96-16866,
AST98-20314 and AST02-06010.  PLS thanks the John Simon Guggenheim
Memorial Foundation its support and the Institute for Advanced Study for its
hopsitality.
\clearpage
\begin{figure}[h]
\vspace{8.0 truein}
\includegraphics{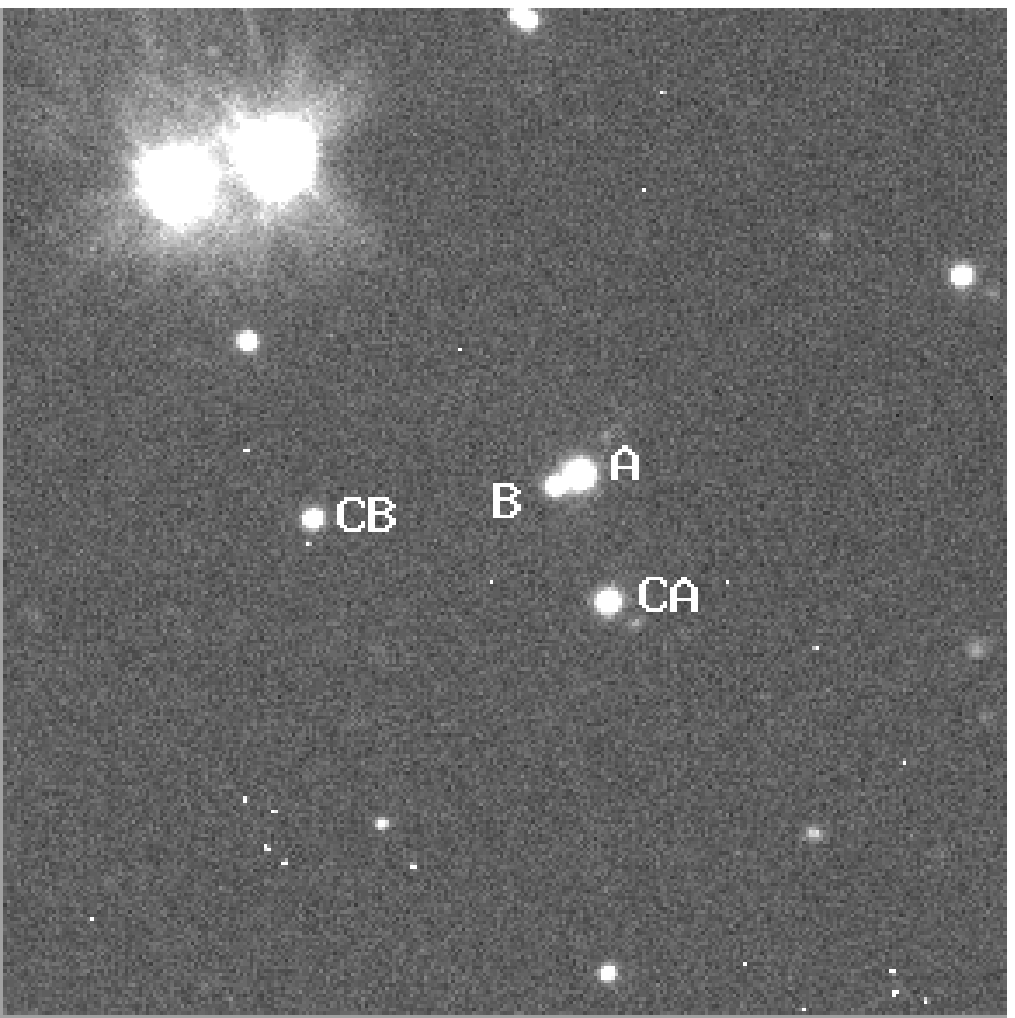}
\caption
{A $120'' \times 120''$ $V$ filter direct image of HE1104$-$1805 showing
QSO components A and B and comparison stars CA ($V = 17.65 \pm 0.02$) 
and CB ($V = 18.43 \pm 0.03$).  North is up and East to the left.
}
\end{figure}

\clearpage
\begin{figure}[h]
\vspace{8.0 truein}
\includegraphics{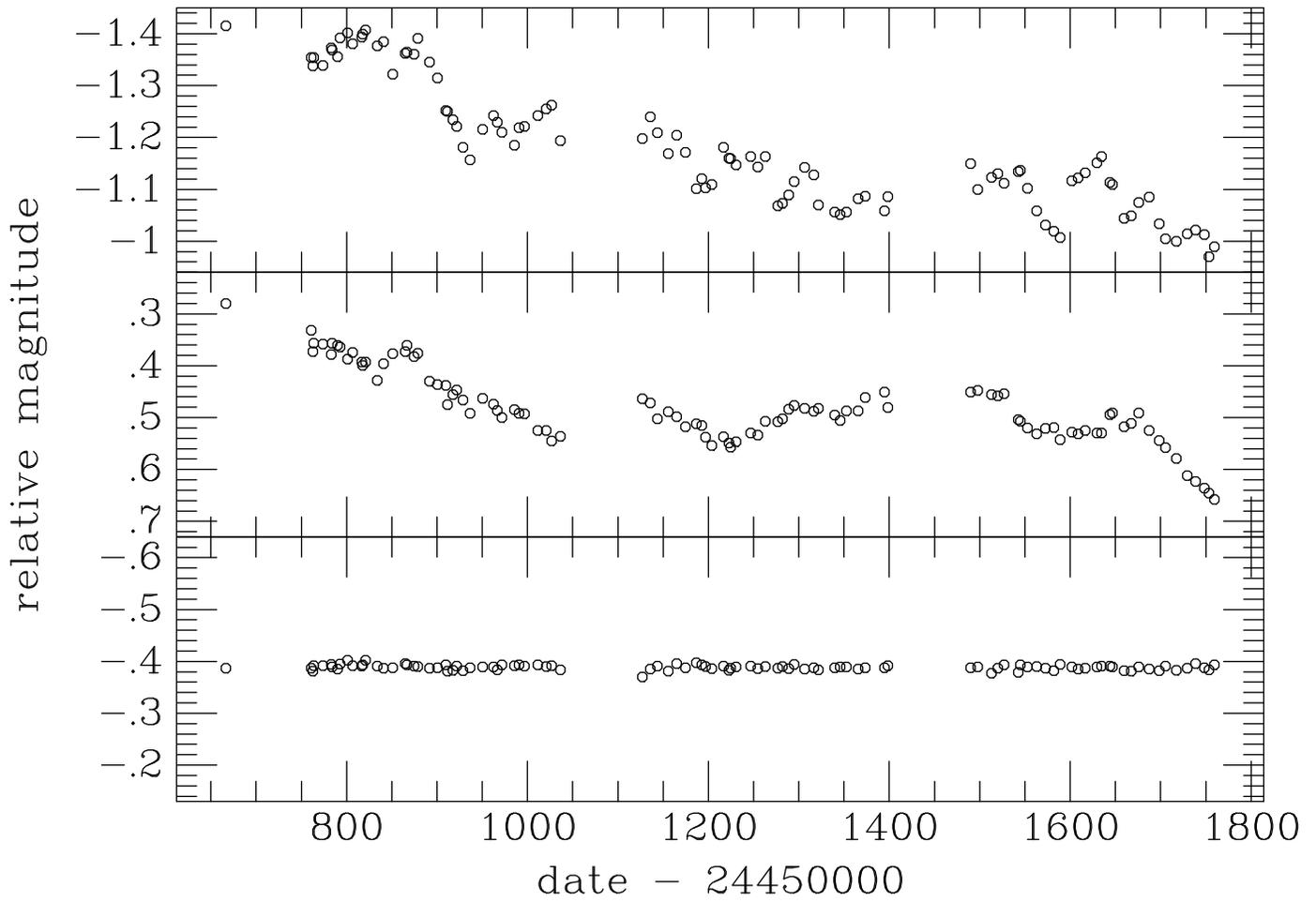}
\caption
{Lightcurves for QSO  components $A$ (top) and $B$ (middle)
and for comparison star $CA$ (bottom).
}
\end{figure}

\clearpage

\begin{figure}[h]
\vspace{8.0 truein}
\includegraphics{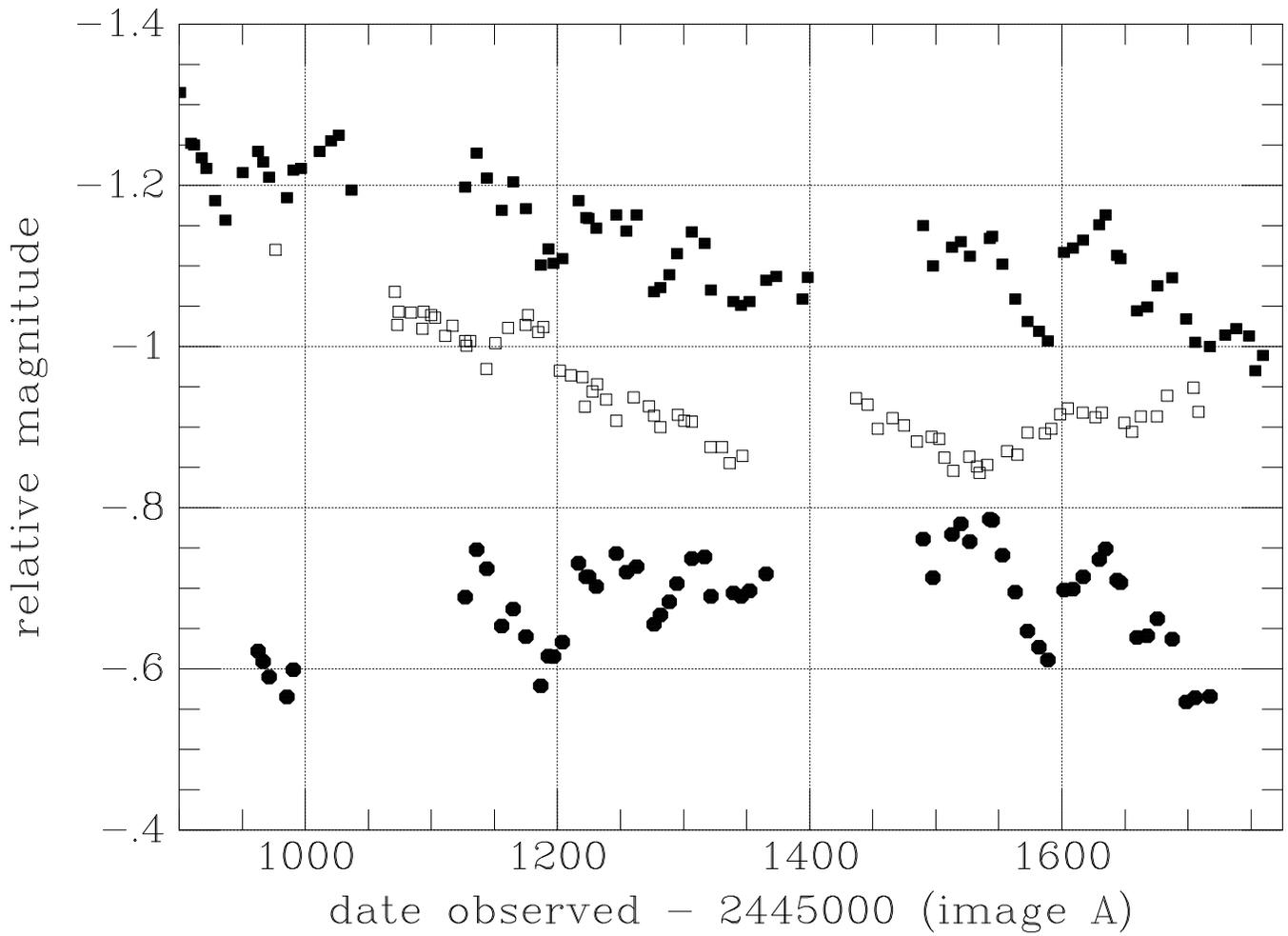}
\caption
{filled squares $\blacksquare$ -- lightcurve for QSO component $A$;
open squares $\square$ -- $B$ lightcurve shifted horizontally 310 days
to a later date and brighter by 1.4 mag; filled circles $\bullet$ --
the difference between the $A$ lightcurve and interpolation on upon
the shifted $B$ lightcurve, shifted fainter by 0.9 mag.
}
\end{figure}
\clearpage

\begin{figure}[h]
\vspace{8.0 truein}
\includegraphics{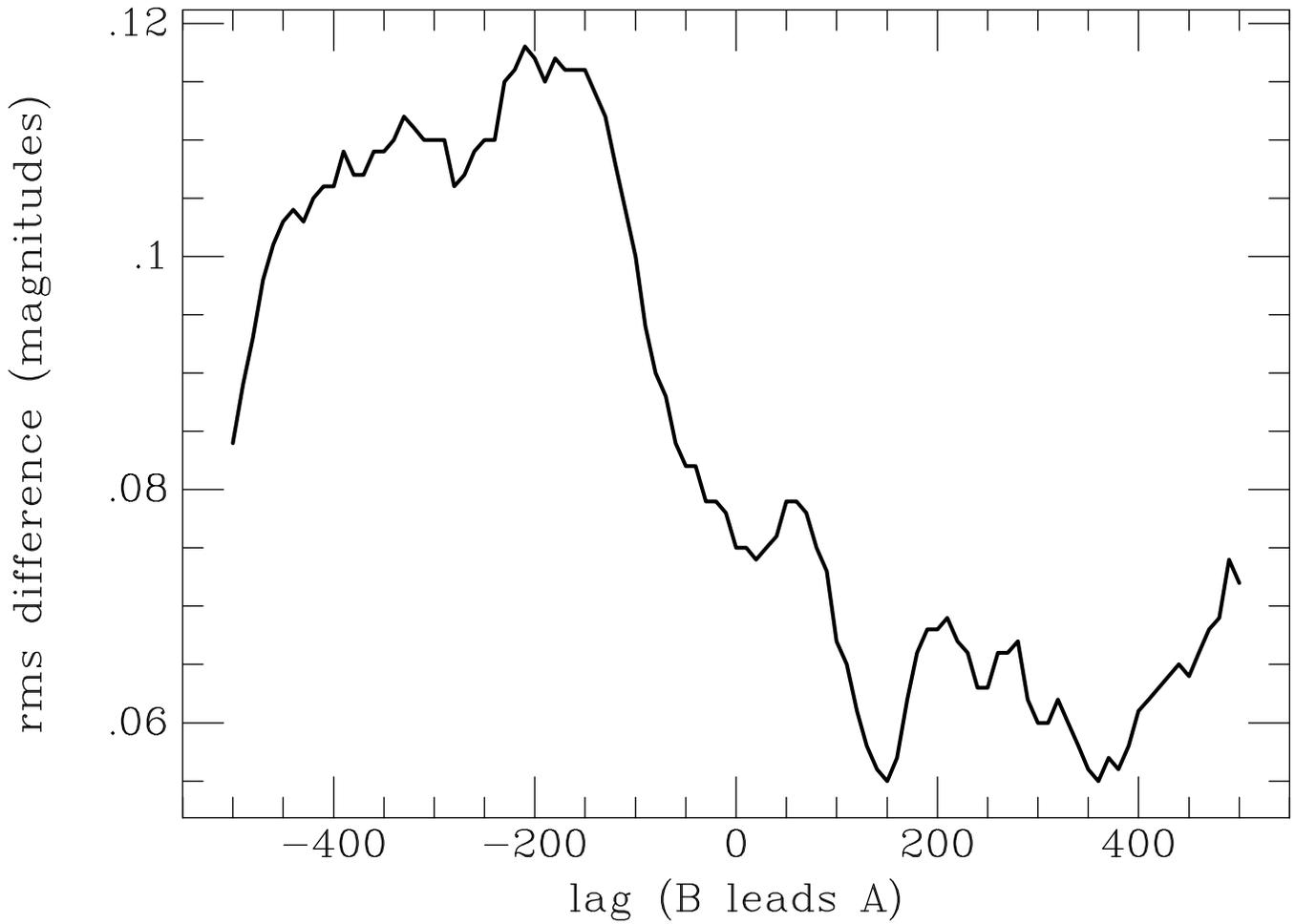}
\caption
{RMS difference between the observed lightcurve for QSO component $A$
and interpolation on $B$ lightcurve shifted to a later date.
}
\end{figure}

\clearpage

\begin{figure}[h]
\vspace{8.0 truein}
\includegraphics{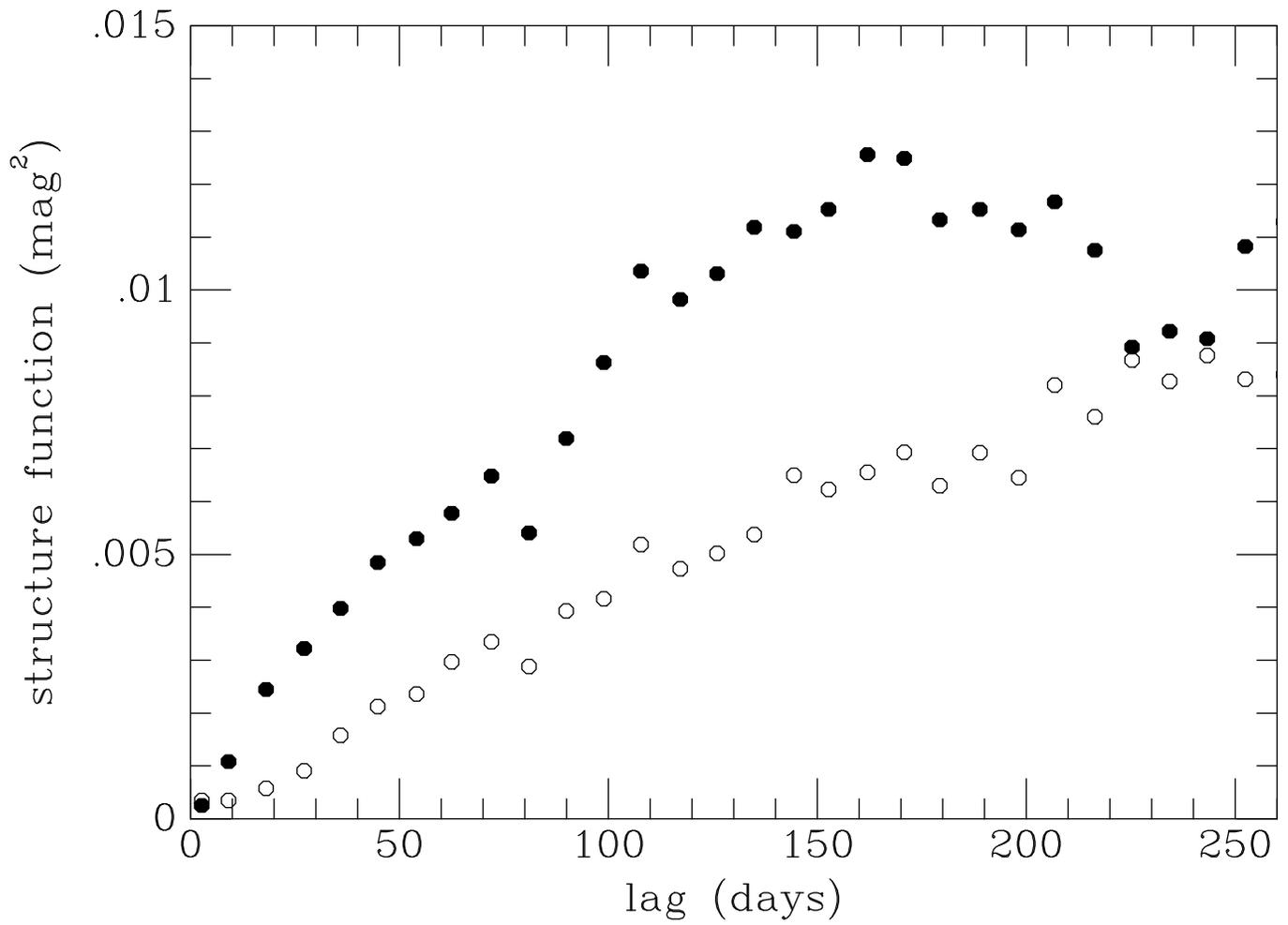}
\caption
{Structure functions for the $A$ ($\bullet$) and $B$ ($\circ$)
lightcurves binned in 9 day intervals.
}
\end{figure}
\clearpage

\begin{figure}[h]
\vspace{8.0 truein}
\includegraphics{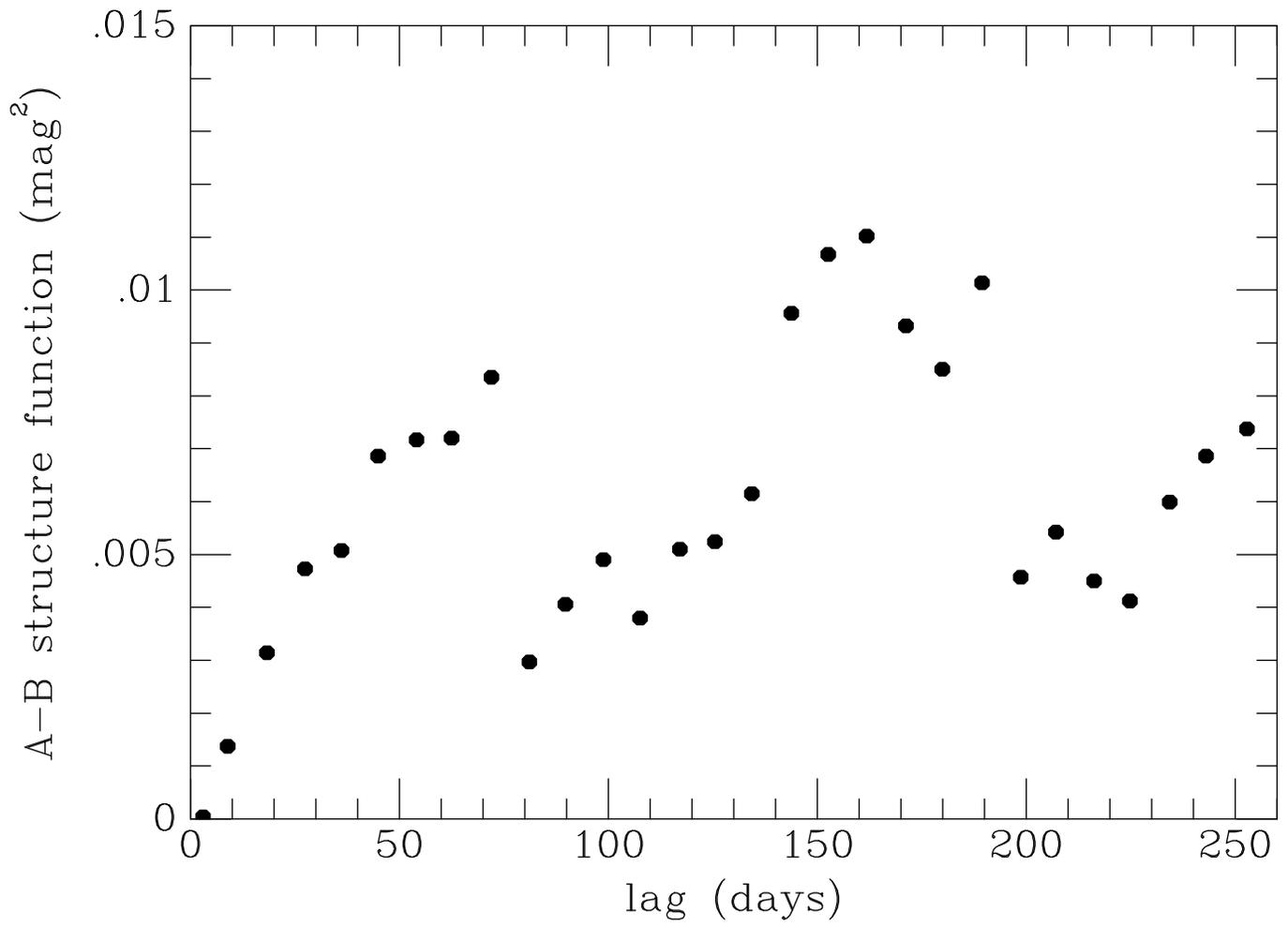}
\caption
{Structure function for the $A-B$ difference lightcurve.  }
\end{figure}
\clearpage

\begin{figure}[h]
\vspace{8.0 truein}
\includegraphics{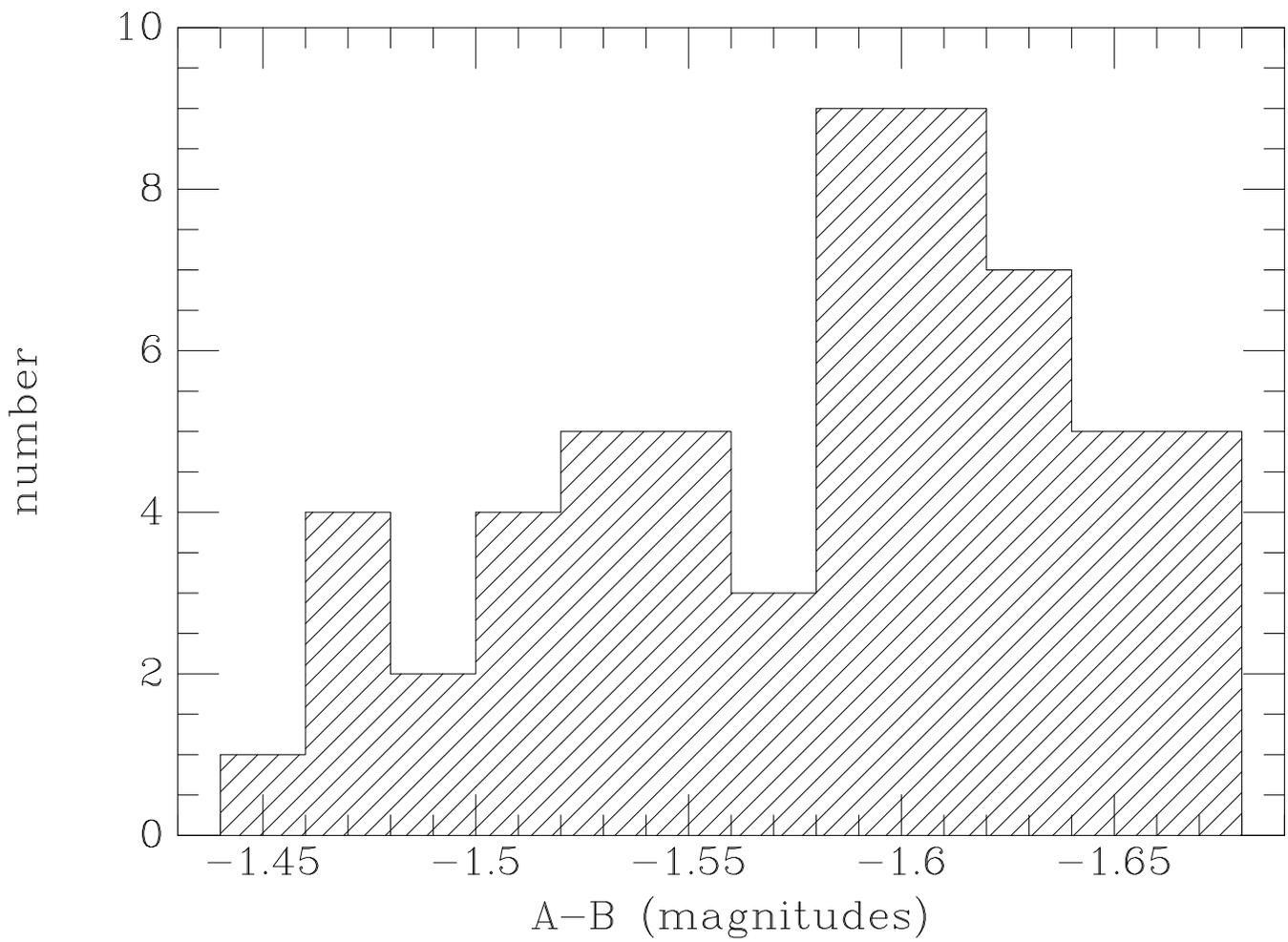}
\caption
{Histogram of $A-B$ magnitude differences}
\end{figure}
\input table1.tex
\input table2.tex

\end{document}

%% file: table1.tex
\clearpage
\begin{deluxetable}{rrrr} 
\tablecolumns{4} 
\tablewidth{0pc} 
\tablecaption{Observations of HE1104-1805 A, B and CA }
\tablehead{ 
\colhead{$
\begin{array}{c}
{\rm date}  \\
({\rm JD} - 2445000)
\end{array}
$}                &
\colhead{$
\begin{array}{c}
m_A \\ 
{\rm (mag)}
\end{array}
$}                & 
\colhead{$
\begin{array}{c}
m_B \\ 
{\rm (mag)}
\end{array}
$}                & 
\colhead{$
\begin{array}{c}
m_{CA} \\ 
{\rm (mag)}
\end{array}
$}                 
}
\startdata

  666.464 & -1.415 & 0.283 & -0.388 \\ 
  666.473 & -1.415 & 0.277 & -0.387 \\
  760.848 & -1.357 & 0.338 & -0.389 \\
  760.856 & -1.351 & 0.327 & -0.385 \\
  762.854 & -1.338 & 0.373 & -0.381 \\
  763.855 & -1.354 & 0.357 & -0.392 \\
  773.847 & -1.338 & 0.362 & -0.393 \\
  773.855 & -1.341 & 0.354 & -0.392 \\
  782.841 & -1.372 & 0.392 & -0.394 \\
  782.850 & -1.375 & 0.365 & -0.393 \\
  783.843 & -1.363 & 0.352 & -0.389 \\
  783.852 & -1.375 & 0.362 & -0.390 \\
  789.850 & -1.356 & 0.361 & -0.385 \\
  792.812 & -1.389 & 0.380 & -0.396 \\
  792.821 & -1.395 & 0.348 & -0.394 \\
  800.843 & -1.405 & 0.385 & -0.403 \\
  800.852 & -1.412 & 0.370 & -0.397 \\
  800.863 & -1.388 & 0.405 & -0.407 \\
  806.852 & -1.375 & 0.380 & -0.390 \\
  806.860 & -1.387 & 0.369 & -0.394 \\
  816.861 & -1.386 & 0.404 & -0.393 \\
  816.872 & -1.402 & 0.381 & -0.388 \\
  817.846 & -1.403 & 0.396 & -0.392 \\
  817.855 & -1.395 & 0.402 & -0.394 \\
  820.868 & -1.407 & 0.393 & -0.402 \\
  833.805 & -1.386 & 0.428 & -0.396 \\
  833.813 & -1.368 & 0.429 & -0.385 \\
  840.791 & -1.386 & 0.402 & -0.390 \\
  840.799 & -1.383 & 0.389 & -0.384 \\
  850.813 & -1.324 & 0.378 & -0.391 \\
  850.821 & -1.320 & 0.376 & -0.385 \\
  864.653 & -1.359 & 0.381 & -0.392 \\
  864.662 & -1.366 & 0.365 & -0.399 \\
  866.779 & -1.370 & 0.370 & -0.392 \\
  866.788 & -1.361 & 0.353 & -0.394 \\
  874.766 & -1.351 & 0.389 & -0.390 \\
  874.775 & -1.371 & 0.376 & -0.391 \\
  878.668 & -1.391 & 0.389 & -0.391 \\
  878.677 & -1.390 & 0.363 & -0.388 \\
  891.741 & -1.346 & 0.430 & -0.386 \\
  891.750 & -1.343 & 0.430 & -0.389 \\
  900.644 & -1.312 & 0.439 & -0.386 \\
  900.652 & -1.317 & 0.433 & -0.390 \\
  909.582 & -1.242 & 0.441 & -0.388 \\
  909.590 & -1.262 & 0.435 & -0.397 \\
  911.773 & -1.247 & 0.459 & -0.376 \\
  911.782 & -1.253 & 0.491 & -0.386 \\
  917.683 & -1.227 & 0.451 & -0.379 \\
  917.691 & -1.240 & 0.461 & -0.387 \\
  921.645 & -1.214 & 0.460 & -0.391 \\
  921.654 & -1.228 & 0.433 & -0.390 \\
  928.629 & -1.181 & 0.459 & -0.382 \\
  928.638 & -1.180 & 0.474 & -0.383 \\
  936.575 & -1.157 & 0.492 & -0.388 \\
  950.574 & -1.204 & 0.464 & -0.381 \\
  950.583 & -1.228 & 0.462 & -0.396 \\
  962.521 & -1.243 & 0.471 & -0.391 \\
  962.529 & -1.240 & 0.476 & -0.387 \\
  966.549 & -1.229 & 0.486 & -0.384 \\
  971.467 & -1.211 & 0.496 & -0.401 \\
  971.477 & -1.209 & 0.502 & -0.389 \\
  971.477 & -1.209 & 0.502 & -0.389 \\
  985.494 & -1.198 & 0.479 & -0.396 \\
  985.503 & -1.172 & 0.491 & -0.388 \\
  990.486 & -1.214 & 0.491 & -0.396 \\
  990.494 & -1.223 & 0.493 & -0.390 \\
  996.469 & -1.218 & 0.484 & -0.387 \\
  996.478 & -1.224 & 0.501 & -0.396 \\
 1011.490 & -1.245 & 0.528 & -0.395 \\
 1011.499 & -1.238 & 0.522 & -0.391 \\
 1020.477 & -1.256 & 0.528 & -0.393 \\
 1020.486 & -1.254 & 0.523 & -0.387 \\
 1026.474 & -1.265 & 0.527 & -0.395 \\
 1026.483 & -1.259 & 0.563 & -0.389 \\
 1036.476 & -1.191 & 0.533 & -0.385 \\
 1036.468 & -1.198 & 0.539 & -0.384 \\
 1126.839 & -1.206 & 0.480 & -0.375 \\
 1126.848 & -1.191 & 0.448 & -0.366 \\
 1135.836 & -1.242 & 0.479 & -0.388 \\
 1135.844 & -1.238 & 0.465 & -0.382 \\
 1143.842 & -1.208 & 0.509 & -0.399 \\
 1143.850 & -1.211 & 0.496 & -0.382 \\
 1155.839 & -1.169 & 0.477 & -0.386 \\
 1155.847 & -1.170 & 0.501 & -0.375 \\
 1164.815 & -1.203 & 0.504 & -0.401 \\
 1164.824 & -1.206 & 0.493 & -0.391 \\
 1174.790 & -1.170 & 0.518 & -0.389 \\
 1174.799 & -1.172 & 0.518 & -0.387 \\
 1186.811 & -1.097 & 0.519 & -0.404 \\
 1186.820 & -1.105 & 0.506 & -0.390 \\
\tablenotemark{a} 1192.759 & -0.852 & 0.523 & -0.389 \\
 1192.768 & -1.121 & 0.515 & -0.393 \\
 1196.783 & -1.096 & 0.545 & -0.391 \\
 1196.791 & -1.110 & 0.530 & -0.388 \\
 1203.783 & -1.106 & 0.558 & -0.389 \\
 1203.792 & -1.112 & 0.549 & -0.382 \\
 1216.777 & -1.182 & 0.540 & -0.391 \\
 1216.785 & -1.180 & 0.534 & -0.391 \\
 1222.769 & -1.158 & 0.557 & -0.383 \\
 1222.777 & -1.162 & 0.541 & -0.384 \\
 1224.755 & -1.147 & 0.554 & -0.379 \\
 1224.764 & -1.171 & 0.559 & -0.396 \\
\tablenotemark{a} 1230.785 & -1.139 & 0.482 & -0.384 \\
 1230.793 & -1.147 & 0.547 & -0.389 \\
 1246.700 & -1.165 & 0.532 & -0.391 \\
 1246.709 & -1.161 & 0.528 & -0.390 \\
 1254.746 & -1.145 & 0.533 & -0.386 \\
 1254.754 & -1.141 & 0.535 & -0.386 \\
 1262.730 & -1.164 & 0.500 & -0.391 \\
 1262.739 & -1.162 & 0.514 & -0.388 \\
 1276.591 & -1.070 & 0.504 & -0.388 \\
 1276.599 & -1.065 & 0.513 & -0.385 \\
 1281.665 & -1.085 & 0.497 & -0.398 \\
 1281.673 & -1.061 & 0.507 & -0.381 \\
 1288.625 & -1.084 & 0.492 & -0.383 \\
 1288.633 & -1.093 & 0.477 & -0.389 \\
 1294.621 & -1.110 & 0.479 & -0.398 \\
 1294.629 & -1.119 & 0.474 & -0.390 \\
 1306.602 & -1.133 & 0.491 & -0.388 \\
 1306.611 & -1.152 & 0.473 & -0.382 \\
 1316.634 & -1.120 & 0.490 & -0.388 \\
 1316.642 & -1.135 & 0.487 & -0.388 \\
 1321.636 & -1.070 & 0.482 & -0.384 \\
 1339.496 & -1.057 & 0.501 & -0.387 \\
 1339.505 & -1.056 & 0.490 & -0.390 \\
 1345.498 & -1.049 & 0.501 & -0.391 \\
 1345.506 & -1.053 & 0.511 & -0.387 \\
 1352.459 & -1.053 & 0.489 & -0.390 \\
 1352.467 & -1.060 & 0.485 & -0.388 \\
 1365.471 & -1.081 & 0.497 & -0.383 \\
 1365.479 & -1.084 & 0.476 & -0.387 \\
 1373.472 & -1.076 & 0.452 & -0.390 \\
 1373.480 & -1.099 & 0.470 & -0.387 \\
 1394.488 & -1.065 & 0.448 & -0.387 \\
 1394.496 & -1.054 & 0.454 & -0.389 \\
 1398.468 & -1.088 & 0.499 & -0.396 \\
 1398.477 & -1.083 & 0.464 & -0.388 \\
 1489.848 & -1.143 & 0.449 & -0.388 \\
 1489.857 & -1.156 & 0.452 & -0.388 \\
 1497.850 & -1.100 & 0.448 & -0.389 \\
 1512.829 & -1.129 & 0.448 & -0.373 \\
 1512.838 & -1.118 & 0.464 & -0.382 \\
 1519.829 & -1.133 & 0.455 & -0.388 \\
 1519.837 & -1.126 & 0.460 & -0.386 \\
 1526.838 & -1.116 & 0.435 & -0.392 \\
 1526.846 & -1.109 & 0.473 & -0.394 \\
 1542.853 & -1.134 & 0.504 & -0.379 \\
 1544.819 & -1.135 & 0.517 & -0.388 \\
 1544.828 & -1.138 & 0.496 & -0.398 \\
 1552.838 & -1.101 & 0.526 & -0.388 \\
 1552.847 & -1.103 & 0.514 & -0.389 \\
 1562.843 & -1.065 & 0.536 & -0.387 \\
 1562.851 & -1.054 & 0.526 & -0.393 \\
 1572.817 & -1.032 & 0.516 & -0.392 \\
 1572.826 & -1.030 & 0.527 & -0.382 \\
 1581.779 & -1.018 & 0.531 & -0.387 \\
 1581.787 & -1.020 & 0.507 & -0.377 \\
 1588.770 & -1.010 & 0.545 & -0.395 \\
 1588.782 & -1.003 & 0.541 & -0.393 \\
 1601.780 & -1.118 & 0.538 & -0.387 \\
 1601.788 & -1.115 & 0.518 & -0.391 \\
 1608.769 & -1.119 & 0.528 & -0.385 \\
 1608.778 & -1.126 & 0.535 & -0.385 \\
 1616.752 & -1.129 & 0.519 & -0.390 \\
 1616.760 & -1.135 & 0.531 & -0.385 \\
 1629.673 & -1.142 & 0.533 & -0.386 \\
 1629.682 & -1.160 & 0.527 & -0.392 \\
 1634.668 & -1.170 & 0.520 & -0.393 \\
 1634.677 & -1.156 & 0.541 & -0.389 \\
 1643.640 & -1.096 & 0.495 & -0.387 \\
 1643.649 & -1.130 & 0.493 & -0.394 \\
 1646.583 & -1.108 & 0.506 & -0.385 \\
 1646.591 & -1.110 & 0.477 & -0.393 \\
 1659.625 & -1.037 & 0.526 & -0.376 \\
 1659.634 & -1.052 & 0.511 & -0.388 \\
 1667.599 & -1.049 & 0.512 & -0.375 \\
 1667.607 & -1.049 & 0.510 & -0.387 \\
 1675.566 & -1.070 & 0.481 & -0.388 \\
 1675.575 & -1.080 & 0.501 & -0.389 \\
 1687.534 & -1.088 & 0.522 & -0.387 \\
 1687.542 & -1.082 & 0.529 & -0.383 \\
 1698.517 & -1.033 & 0.538 & -0.383 \\
 1698.525 & -1.035 & 0.549 & -0.381 \\
 1705.539 & -1.001 & 0.567 & -0.389 \\
 1705.547 & -1.009 & 0.548 & -0.392 \\
 1717.464 & -1.000 & 0.574 & -0.390 \\
 1717.483 & -1.000 & 0.571 & -0.384 \\
 1717.491 & -0.999 & 0.592 & -0.374 \\
 1729.488 & -1.018 & 0.600 & -0.387 \\
 1729.496 & -1.010 & 0.623 & -0.387 \\
 1738.481 & -1.026 & 0.610 & -0.400 \\
 1738.489 & -1.018 & 0.636 & -0.393 \\
 1748.489 & -1.013 & 0.636 & -0.388 \\
 1753.476 & -0.972 & 0.645 & -0.391 \\
 1753.484 & -0.969 & 0.647 & -0.378 \\
\tablenotemark{a} 1759.454 & -1.005 & 0.565 & -0.388 \\
 1759.463 & -0.989 & 0.658 & -0.393 \\
\enddata 
\tablenotetext{a}{Observation discarded}
\end{deluxetable} 

%% file: table2.tex
\clearpage
\begin{deluxetable}{rrrr} 
\tablecolumns{4} 
\tablewidth{0pc} 
\tablecaption{Lensing Parameters for Images A and B}
\tablehead{ 
\colhead{Image}   &  \colhead{$\kappa_{tot}$} & 
\colhead{$\gamma$} & \colhead{$\mu$}
}\startdata
  A &   0.639 &   0.521 &  $-$7.08  \\
  B &   0.335 &   0.215 &  +2.53  \\
\enddata 
\end{deluxetable}